\documentclass{PoS}
\usepackage{amsmath}
\usepackage{amssymb}
\usepackage{leftidx}
\usepackage{graphicx}
\usepackage[utf8]{inputenc}
\usepackage{newunicodechar}

\inputencoding{utf8}
\newunicodechar{γ}{$\gamma$}
\newunicodechar{ρ}{$\rho$}
\newunicodechar{π}{$\pi$}
\newunicodechar{Δ}{$\Delta$}
\newunicodechar{η}{$\eta$}
\newunicodechar{ψ}{$\psi$}
\newunicodechar{→}{$\to$}

\newcommand{\I}{\mathrm{i}}
\newcommand{\ampV}{{\cal V}_{\pi\gamma \to \pi\pi}}

\newcommand{\formF}{F}
\newcommand{\vecQ}{\vec{q}}
\newcommand{\vecP}{\vec{P}}
\newcommand{\vecp}{\vec{p}_{\pi}}

\newcommand{\tSrc}{t_{\pi\pi}}
\newcommand{\tSnk}{t_{\pi}}
\newcommand{\tCurr}{t_{J}}
\newcommand{\DeltaT}{\Delta t}

\title{Calculating the $\rho$ radiative decay width with lattice QCD}

\ShortTitle{Calculating the $\rho$ radiative decay width with lattice QCD}

\author{\speaker{Luka Leskovec}$^{\,\,a}$,
		Constantia Alexandrou$^{b,c}$,
        Stefan Meinel$^{d,e}$,
        John W. Negele$^{f}$,
        Srijit Paul$^{c}$,
        Marcus Petschlies$^{g}$,
        Andrew Pochinsky$^{f}$, 
        Gumaro Rendon$^{d}$, 
        Sergey Syritsyn$^{h,e}$\\

		$^{a}$Thomas Jefferson National Accelerator Facility, 12000 Jefferson Avenue\\
        Newport News, Virginia 23606, USA\\
    $^{b}$Department of Physics, University of Cyprus, P.O. Box 20537, 1678 Nicosia, Cyprus\\
    $^{c}$Computation-based Science and Technology Research Center, Cyprus Institute, 20 Kavafi Street, 2121 Nicosia, Cyprus\\
    $^{d}$Department of Physics, University of Arizona, Tucson, Arizona 85721, USA\\
    $^{e}$RIKEN BNL Research Center, Brookhaven National Laboratory, Upton, New York 11973, USA\\
		$^{f}$Center for Theoretical Physics, Laboratory for Nuclear Science and Department of Physics,Massachusetts Institute of Technology, Cambridge, Massachusetts 02139, USA
		$^{g}$Helmholtz-Institut f{\"u}r Strahlen- und Kernphysik, Rheinische Friedrich-Wilhelms-Universit{\"a}t Bonn, Nu{\ss}allee 14-16, D-53115 Bonn, Germany\\
		$^{h}$Department of Physics and Astronomy, Stony Brook University, Stony Brook, New York 11794, USA\\
	}

\abstract{
We present the results of our lattice QCD study of the $\pi\gamma\to\pi\pi$ process, where the $\rho$ resonance appears as an enhancement in the transition amplitude. We use $N_f=2+1$ clover fermions on a lattice of $L=3.6$ fm and a pion mass of $320$ MeV. Using a combination of forward, stochastic, and sequential propagators, we calculate the two-point and three-point functions that allow us to determine the $\pi\gamma\to\pi\pi$ matrix elements for several values of the invariant mass $s$ and momentum transfer $q^2$. To fit the $q^2$ and $s$ dependence of the $\pi\gamma\to\pi\pi$ amplitude, we explore a set of general parametrizations based on a Taylor expansion. By analytic continuation to the complex pole corresponding to the $\rho$ resonance, we determine the resonant form factors and calculate the radiative decay width of the $\rho$.
}

\FullConference{The 36th Annual International Symposium on Lattice Field Theory - LATTICE2018\\
		22-28 July, 2018\\
		Michigan State University, East Lansing, Michigan, USA.}

\begin{document}

\section{Introduction}
The spectrum of hadrons provides a unique connection between QCD and experiment. Many experiments in hadron spectroscopy have utilized elastic and charge-exchange $\pi N$ scattering,
but certain resonances are difficult to discern from this production mechanism, which leads to the ``missing resonance problem'' in the spectrum of baryons \cite{Koniuk:1979vw}. Production processes other than $\pi N$ scattering allow us to better probe the spectrum of hadrons and increase our understanding of the underlying physics \cite{Ronchen:2015vfa}. One possible alternative production process that is being utilized at modern-day facilities like JLab, ELSA, and MAMI is photoproduction. There, a photon is absorbed on a stable hadron, like a proton, which is then excited into a resonance that decays to various lighter hadrons. Recent theoretical developments \cite{Briceno:2014uqa,Briceno:2015csa} now allow lattice QCD calculations of the photoproduction of two-body resonances. However, as the baryon sector is very challenging, it is helpful to first investigate the photoproduction of the simplest known resonance, the $\rho$ meson. Here, we give an overview of our lattice QCD calculation of this process \cite{Alexandrou:2018jbt}.

\section{Description of the photoproduction}
Because the $\rho$ is a resonance and thus not stable under the strong interaction, the photoproduction process $\pi\gamma\to \rho$ is obtained by analytical continuation of the more general process $\pi\gamma\to\pi\pi$ to the $\rho$ pole. Here the $\pi\pi$ state is in $P$-wave, has isospin $I=1$ and $J^{PC}=1^{--}$. The process $\pi\gamma\to\pi\pi$ is described by the infinite-volume matrix element $\langle \pi \pi |J_\mu |\pi\rangle$, which has the Lorentz decomposition
\begin{align}
\label{eq:LLdecomp}
\langle \pi\pi | J^\mu(0) | \pi \rangle = \frac{2\I \ampV(q^2,s)}{m_\pi} \epsilon^{\nu\mu\alpha\beta} \epsilon_{\nu}(P,m) (p_\pi)_\alpha P_\beta.
\end{align}
Above, $\ampV$ is the transition amplitude, which depends on the momentum transfer $q^2 = (p_\pi - P)^2$, $P$ is the $\pi\pi$-state four-momentum vector, $p_\pi$ is the $\pi$-state four-momentum vector and $\epsilon_{\nu}(P,m)$ is the $\pi\pi$-state polarization vector. Further details on the states, their normalization and specific implementations can be found in Ref.~\cite{Alexandrou:2018jbt}. Near the $\rho$ pole, $s_P=m_R^2 + \I m_R \Gamma_R$, the transition amplitude $\ampV$ takes the form
\begin{align}
\label{eq:ampV}
\ampV(0, s) \sim \frac{G_{\rho\pi\pi} \:G_{\rho\pi\gamma}}{s_P - s},
\end{align}
where $G_{\rho\pi\gamma}$ is the $\rho$ resonance photocoupling and $G_{\rho\pi\pi}$ is the coupling of the $\rho$ to the $\pi\pi$ channel. Making use of Watson's theorem, the transition amplitude can be reformulated as
\begin{align}
\label{eq:V_BW_d}
\ampV(q^2,s) & =  \sqrt{\frac{16 \pi}{k\Gamma(s)}} \frac{\formF(q^2,s) }{\cot{\delta(s) - \I}},
\end{align}
where $\delta$ is the phase shift describing the strong $\pi\pi$ scattering. We investigate the two Breit-Wigner forms described in Ref.~\cite{Alexandrou:2017mpi}, where BWI is the classic $P$-wave Breit-Wigner and BWII is modified by a Blatt-Weisskopf barrier factor. For complete generality, the form factor $\formF(q^2,s)$ still carries an $s$ dependence, but it no longer has a pole in the $s$ plane within the energy region of interest, below the $K\bar{K}$ threshold.

\section{Finite-volume matrix elements with lattice QCD}
\begin{figure}[htb!]
  \centering
  \includegraphics[width=0.29\textwidth]{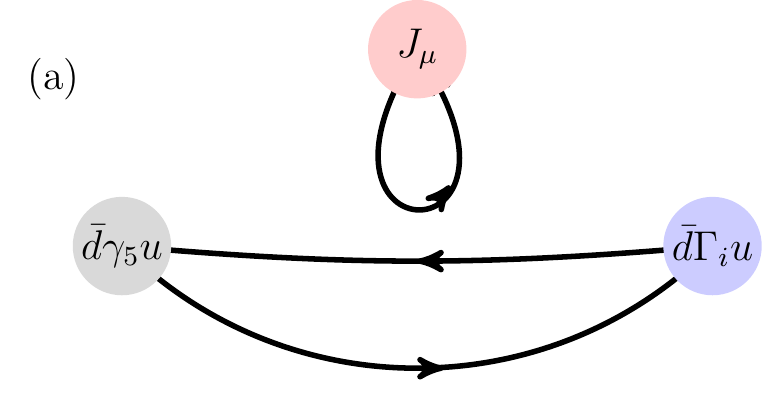}%
  \includegraphics[width=0.29\textwidth]{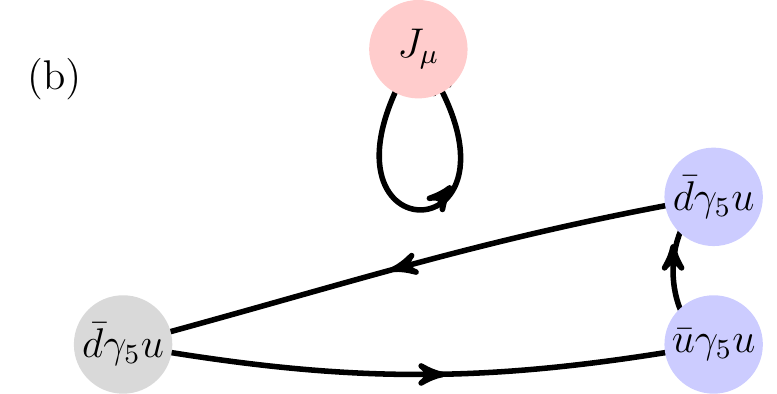}%
  \includegraphics[width=0.29\textwidth]{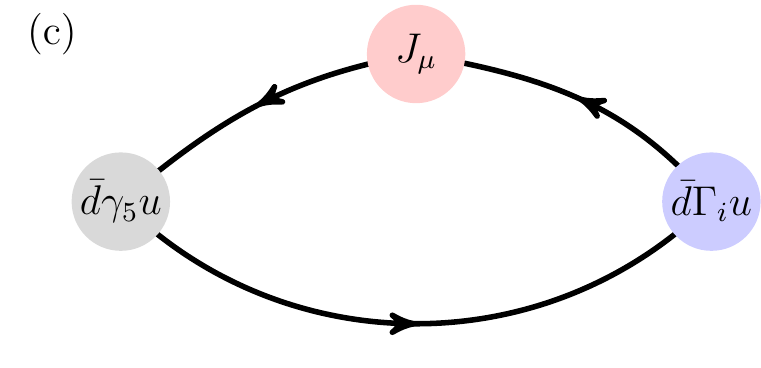}
  \includegraphics[width=0.29\textwidth]{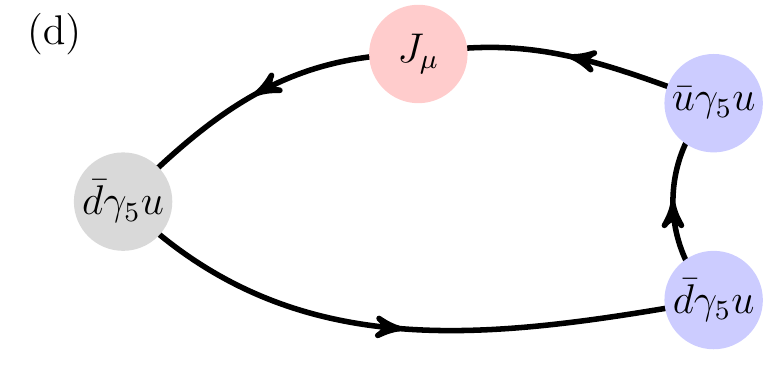}%
  \includegraphics[width=0.29\textwidth]{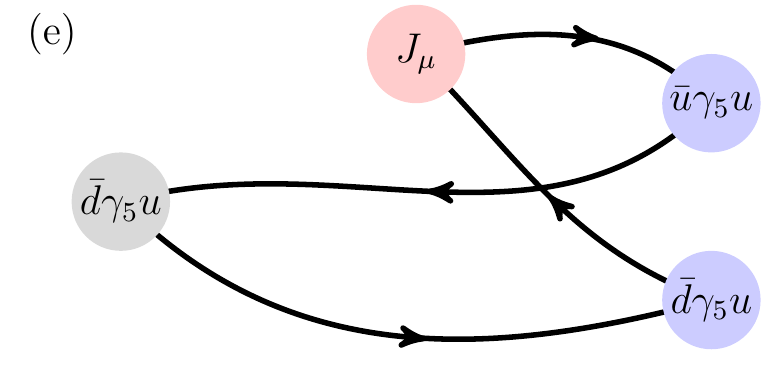}%
  \includegraphics[width=0.29\textwidth]{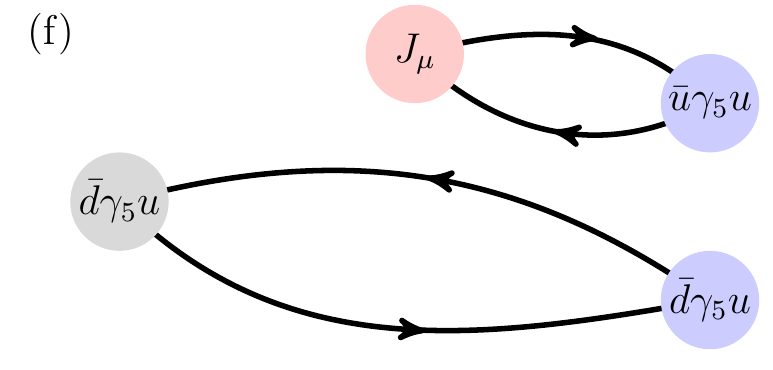}
  \caption{
  \label{fig:3pt} The different topologies of Wick contractions that make up the three-point function $C_{3,\;\mu,i}^{\vecp,\,\vecP,\,\Lambda,\,r}(\tSnk,\tCurr,\tSrc)$. The current-disconnected diagrams a) and b) are not included in the construction of the three-point functions.
  }
\end{figure}

We use a single gauge-field ensemble with $N_f=2+1$ clover-Wilson fermions on a $N_s^3 \times N_t = 32^3\times96$ lattice. The lattice spacing determined from the $\Upsilon(1S)$-$\Upsilon(2S)$ splitting with NRQCD is $a=0.11403(77)$ fm. Our light-quark masses correspond to a pion mass of $am_\pi=0.18295(36)$. Our calculation of the $\rho$ resonance parameters using the L\"uscher method in several moving frames and irreducible representations can be found in Ref.~\cite{Alexandrou:2017mpi}.
To determine the finite-volume matrix elements $\langle \pi,\vecp | J_\mu(0,\vecQ)|n,\vecP,\,\Lambda,\,r\rangle_{FV}$, we calculate the three-point functions
\begin{align}
\label{eq:C3_decomp}
& C_{3,\;\mu,i}^{\vecp,\,\vecP,\,\Lambda,\,r}(\tSnk,\tCurr,\tSrc) = \langle  O_{\pi}^{\vecp}(\tSnk)  J_\mu(\tCurr,\vecQ) O_{i}^{n,\vecP,\,\Lambda,\,r}(\tSrc) \rangle \cr
&= \sum_{n} Z_\pi^{\vecp}\: Z_i^{n,\,\vecP,\,\Lambda\,\dag}\: \langle \pi,\vecp | J_\mu(0,\vecQ)|n,\vecP,\,\Lambda,\,r\rangle_{FV}
 \times \frac{e^{-E_\pi^{\vecp} (\tSnk - \tCurr)} e^{-E_n^{\vecP, \Lambda} (\tCurr - \tSrc)} }{2E_n^{\vecP, \Lambda } 2E_\pi^{\vecp}},
\end{align}
where $i$ indexes the single- and two-hadron interpolating operators used to construct the $\pi\pi$ state with $J^{PC}=1^{--}$ and $I=1$ \cite{Alexandrou:2018jbt}. The initial two-pion state has momentum $\vecP$, is in the $r$th row of the finite-volume irreducible representation $\Lambda$, and has energy $E_{n}^{\vecP,\Lambda}$ and overlap factors $Z_i^{n,\vecP,\Lambda}$. The final state has momentum $\vecp$ and energy $E_\pi^{\vecp}$ with the overlap factor $Z_\pi^{\vecp}$. The electromagnetic current insertion $J_\mu = Z_V (\frac{1}{3}\bar{u}\gamma_\mu u -\frac{2}{3} \bar{d}\gamma_\mu d)$ is renormalized as described in Ref.~\cite{Green:2017keo}. The multi-hadron three-point functions are a linear combination of the Wick contractions presented in Fig.~\ref{fig:3pt} and are calculated with forward, sequential and stochastic propagators. We have checked that the the current-disconnected diagrams are statistically compatible with $0$ and omit them from our construction. 
The three-point functions as written in Eq.~(\ref{eq:C3_decomp}) are still a sum over all states in the $\pi\pi$ channel and need to be projected to the finite-volume states which have well-defined invariant mass $s$ and momentum transfer $q^2$. We achieve this by making use of the generalized eigenvectors $v_i^{n\,\vecP, \Lambda}(t_0)$ from the variational analysis in Ref.~\cite{Alexandrou:2017mpi}, and construct the optimized three-point functions \cite{Dudek:2009kk,Becirevic:2014rda,Shultz:2015pfa}:
\begin{align}
\label{eq:omega}
&\Omega_{3,\;\mu,\, n}^{\vecp,\,\vecP,\,\Lambda,\,r}(\tSnk,\tCurr,\tSrc, t_0) =  v_i^{n\,\vecP, \Lambda}(t_0)\:C_{3,\;\mu,i}^{\vecp,\,\vecP,\,\Lambda,\,r}(\tSnk,\tCurr,\tSrc) \cr
&= \sqrt{2E_n^{\vecP, \Lambda}}\, e^{E_n^{\vecP, \Lambda} t_0 / 2} \, Z_\pi^{\vecp} \, \langle \pi,\vecp |J_\mu(0,\vecQ) | n, \vecP, \Lambda,r \rangle_{FV}\times\frac{ e^{-E_\pi^{\vecp} (\tSnk - \tCurr)} e^{-E_n^{\vecP, \Lambda} (\tCurr - \tSrc)}}{2E_n^{\vecP, \Lambda} 2E_{\pi}^{\vecp}}.
\end{align}
Here, the sum over all states $n$ in the irreducible representation is reduced to a single term which enables us to use all individual excited states within the energy region of interest to determine the matrix element at as many points as allowed by the quark masses and volume of the gauge ensemble.

The matrix elements $\langle \pi,\vecp | J_\mu(0,\vecQ)|n,\vecP,\,\Lambda,\,r\rangle_{FV}$ are determined using the ratio
\begin{align}
\label{eq:ratio}
& R_{\mu,\, n}^{\vecp,\,\vecP,\,\Lambda,\,r}(\tSnk,\tCurr,\tSrc) = \frac{|\langle \pi,\vecp | J_\mu(0,\vecQ)|n,\vecP,\,\Lambda,\,r\rangle_{FV}|^2}{4 E_n^{\vecP, \Lambda} E_\pi^{\vecp}} \\
& = \frac{\Omega_{3,\;\mu,\, n}^{\vecp,\,\vecP,\,\Lambda,\,r}(\tSnk,\tCurr,\tSrc, t_0)\: \Omega_{3,\;\mu,\, n}^{\vecp,\,\vecP,\,\Lambda,\,r\,\dag}(\tSnk, t^\prime,\tSrc, t_0)}{C_{\pi}^{\vecp}(\DeltaT)\:\lambda_n^{\vecP, \Lambda}(\DeltaT,t_0)},
\end{align}
which depends on the optimized three-point functions as well as the two point functions.

\section{Mapping to infinite volume}
To obtain the infinite-volume matrix elements we follow the Brice\~no-Hansen-Walker-Loud approach \cite{Briceno:2014uqa,Briceno:2015csa}, which is a generalization of the seminal work of Lellouch and L\"uscher \cite{Lellouch:2000pv} and allows us to utilize any gauge ensemble to study general transitions between single-hadron and two-hadron states. The mapping in the case of $\pi\gamma\to\pi\pi$ looks like
\begin{align}
\label{eq:LLmap}
& \frac{|\langle \pi, \vec{p}_\pi |J_\mu(0) |  s,q^2 ; \vecP,\,\Lambda,\,r \rangle_{IV}|^2}{|\langle \pi,\vecp | J_\mu(0,\vecQ)|n,\vecP,\,\Lambda,\,r\rangle_{FV}|^2}  = \frac{1 }{2E_n^{\vecP,\Lambda}}    \frac{16 \pi \sqrt{s_n^{\vecP,\Lambda}}}{k^{\vecP,\Lambda}_n} \left( \frac{\partial \delta}{\partial E} + \frac{\partial \phi^{\vecP,\Lambda}}{\partial E} \right) \bigg|_{E=E_n^{\vecP,\Lambda}}, \qquad
\end{align}
where $\phi^{\vecP,\Lambda}$ are the combinations of the finite-volume quantization condition (cf.~Eq.~49 of Ref.~\cite{Alexandrou:2017mpi}) and $\delta$ is the scattering phase shift as determined in Ref.~\cite{Alexandrou:2017mpi}. We make use of two different parametrizations of the scattering phase shift and perform the infinite-volume mapping for both. The systematic uncertainties associated with the different choices in fitting the energies and matrix elements are propagated to the final results.

\section{The $\pi\gamma \to \pi\pi$ amplitude}
\begin{center}
\begin{figure}[htb]
\centering
  \includegraphics[width=0.67\textwidth]{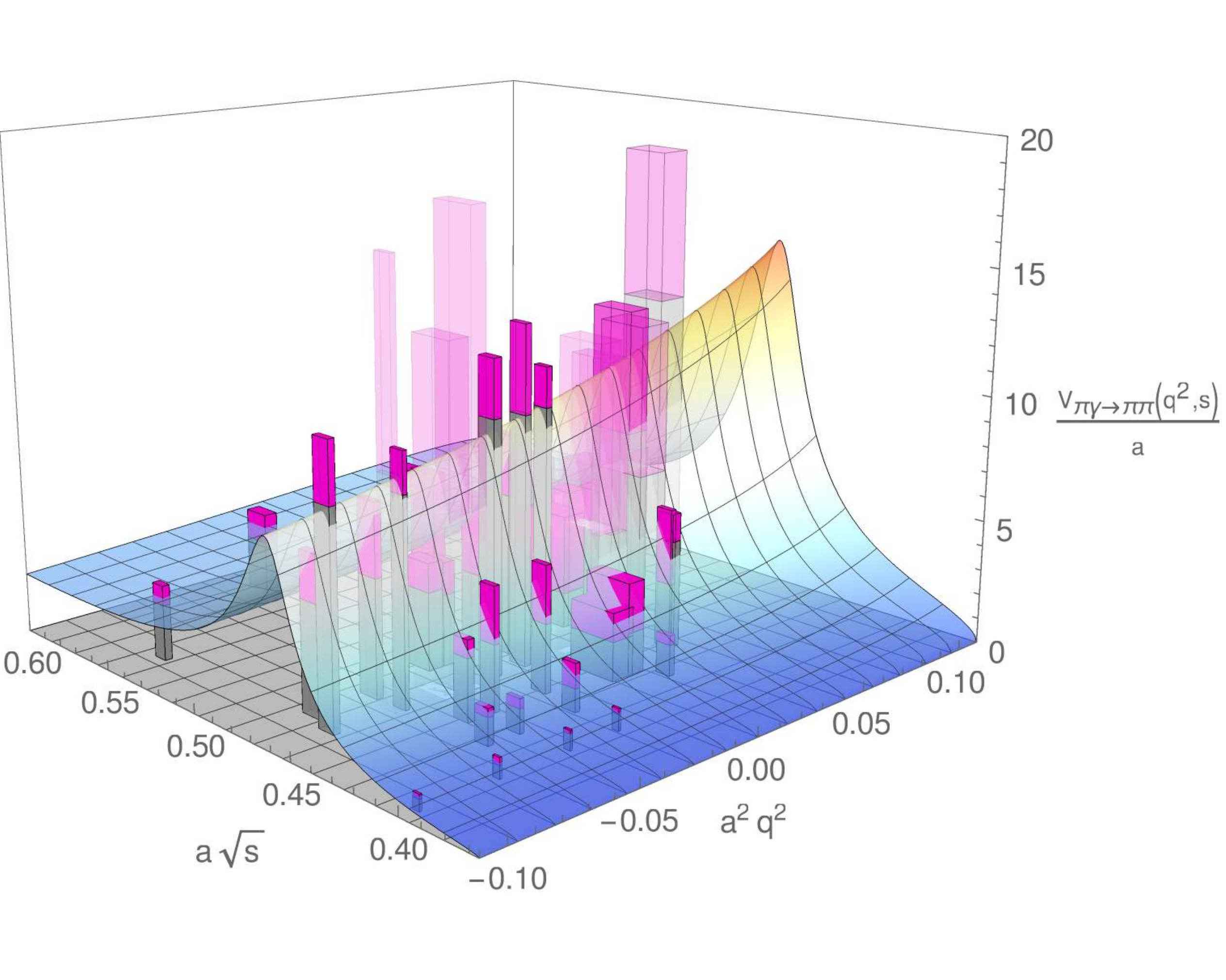}  
  \caption{\label{fig:V_3d} A three-dimensional plot of $\ampV$. The surface shows the central value of the nominal fit function (``BWII F1 K2'').  The lattice results are shown as the vertical bars, with the widths and depths corresponding to the uncertainties. The magenta sections in the vertical direction cover the range from $\ampV-\sigma_{\ampV}$ to $\ampV+\sigma_{\ampV}$. Data points with larger uncertainties have reduced opacity.}
\end{figure}
\end{center}

We build the transition amplitude $\ampV$ (see Eq.~\ref{eq:V_BW_d}) from phase shifts BW I and BW II and parametrize the transition form factor $\formF$ with a two-dimensional Taylor series:
\begin{align}
\formF(q^2,s) = \frac{1}{1 - \frac{q^2}{m_P^2}} \sum_{n,m} A_{nm} z^n \mathcal{S}^m,
\label{eq:Fseries}
\end{align}
where ${\cal S} = \frac{s - m_R^2}{m_R^2}$, $z = \frac{\sqrt{4m_\pi^2 - q^2} - \sqrt{4m_\pi^2 }}{\sqrt{4m_\pi^2 - q^2} + \sqrt{4m_\pi^2 }}$, and $m_R$ corresponds to the $\rho$ resonance mass. Because we neglect the current-disconnected diagrams, we also set $m_P=m_R$. We investigate three different systematic truncations: {\bf F1)} combined order $K$, $\sum_{n+m \leq K}$,  {\bf F2)} order $N$ in $z$ and combined order $K$, $\sum_{n=0}^{N} \sum_{m=0}^{K-n}$, and {\bf F3)} order $N$ in $z$ and order $M$ in $\mathcal{S}$, $\sum_{n=0}^{N} \sum_{m=0}^{M}$. We present a three-dimensional projection of the transition amplitude $\ampV$ for the chosen parametrization ``BW II F1 K2'' in Fig.~\ref{fig:V_3d}.

\section{Resonant form factor and photocoupling}

\begin{figure}[!t]
\includegraphics[width=0.45\textwidth]{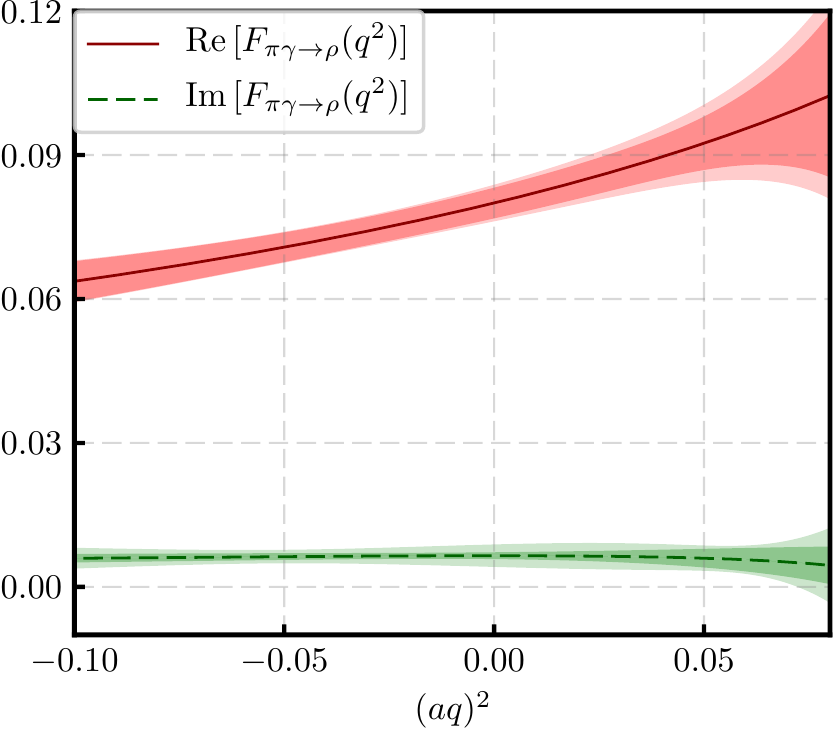}%
\includegraphics[width=0.45\textwidth]{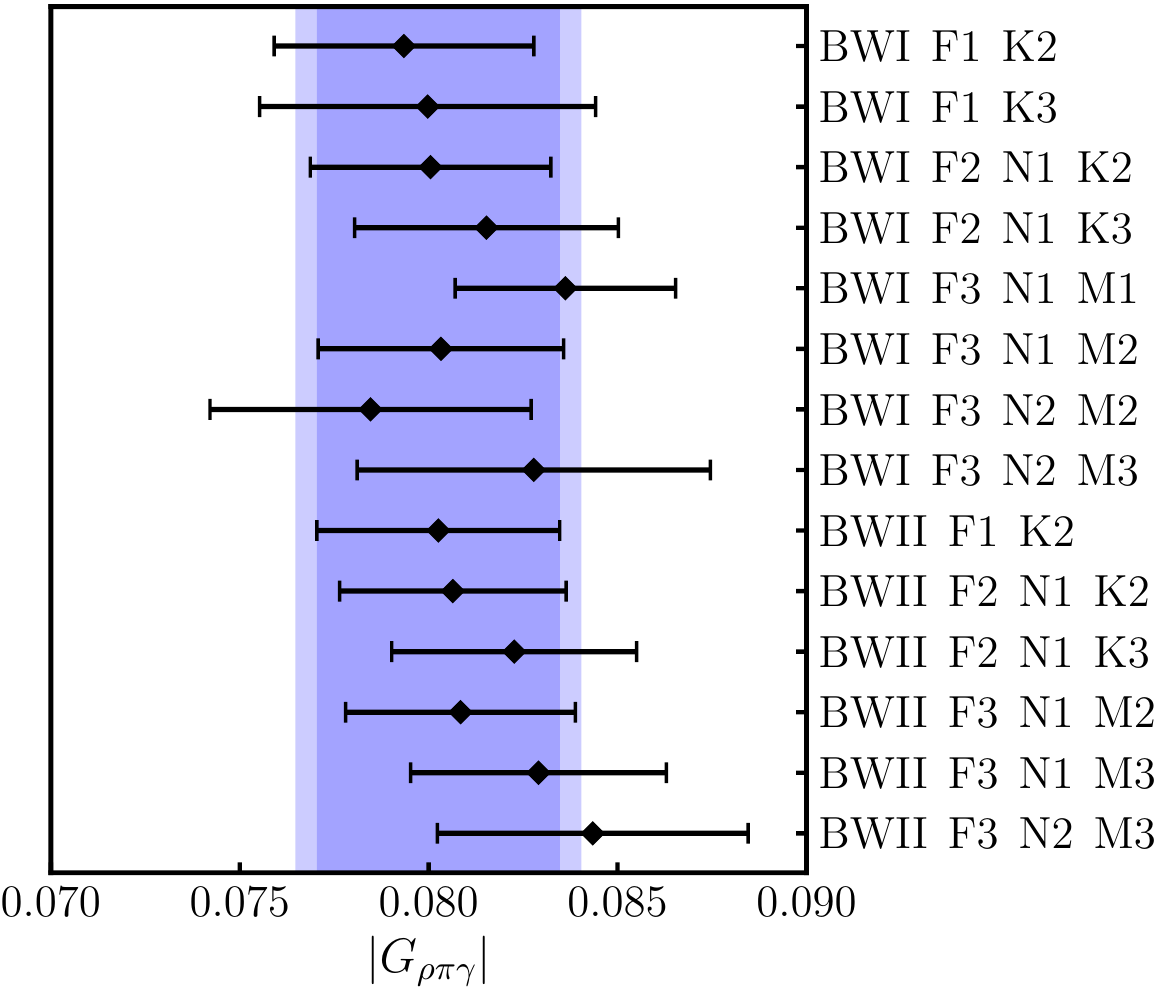}
\caption{\label{fig:PC} Left panel: Resonant transition form factor $\formF_{\pi\gamma\to\rho}$ as a function of momentum transfer. Right panel: Absolute value of the photocoupling for the various parametrizations under consideration.}
\end{figure}

The resonant form factor is determined by analytically continuing the transition form factor $\formF$ to the $\rho$ pole:
\begin{equation}
 \formF_{\pi\gamma\to\rho}(q^2) = \formF(q^2,\, m_R^2 + i m_R \Gamma_R). \label{eq:Fresonant}
\end{equation}
At zero momentum transfer, $q^2=0$, the resonant form factor becomes equal to the photocoupling $G_{\rho\pi\gamma}$, which enters the formula for the $\rho$ radiative decay width $\Gamma(\rho\to\pi\gamma)$:
\begin{align}
&\Gamma(\rho \to \pi \gamma) =\frac{2}{3}\alpha \left(\frac{m_R^2 - m_\pi^2}{2m_R}\right)^3  \frac{|G_{\rho\pi\gamma}|^2}{m_\pi^2}. \label{eq:radwidth}
\end{align}
The left panel of Fig.~\ref{fig:PC} shows the resonant transition form factor as a function of momentum transfer. The inner shaded region corresponds to the statistical and systematical errors and the outer shaded region represents the model uncertainty, evaluated point-by-point as the root-mean-square deviation of the other parametrizations from the nominal parametrization. 
The right panel of Fig.~\ref{fig:PC} shows the photocoupling determined from those parametrizations that lead to a good description of the data without redundant parameters. We find the $\rho$ photocoupling at $m_\pi\approx320$ MeV to be: $|G_{\rho\pi\gamma}| = 0.0802(32)(20)$.
Because our lattice QCD calculation was performed at an unphysical pion mass, our thresholds are much closer to the resonances than in nature and thus our phase space is unphysical. However, if we assume the photocoupling to be pion-mass independent and use the physical values of hadron masses to have the proper thresholds, we find the radiative decay width $\Gamma(\rho\to\pi\gamma) \,=\, 84.2(6.7)(4.3)\, {\rm keV}$. The number in the first bracket represents the combined statistical and systematical uncertainty and the second uncertainty is associated with the parametrization.

\section{Conclusions}

In this talk we presented a $(2+1)$-flavor lattice QCD calculation of the $\rho$ photoproduction process at $m_\pi\approx 320$ MeV. Because at this light-quark mass the $\rho$ is a resonance, we made use of the the Brice\~no-Hansen-Walker-Loud formalism to calculate the general $\pi\gamma\to\pi\pi$ transition amplitude at many discrete points in the kinematic space of $q^2$ and $\sqrt{s}$. Fitting a multitude of parametrizations to the data and analytically continuing them to the $\rho$ pole enabled us to determine the $\rho$ photocoupling and radiative decay width.  Additional computations are needed to perform the continuum extrapolation and the chiral extrapolation to the physical point, where direct comparisons with experimental data are possible.
The methods used in this work are also applicable to other electroweak processes with two hadrons in the final state, such as $B \to \rho (\to \pi \pi) \ell \bar{\nu}$ and $B \to K^* (\to K \pi) \ell^+ \ell^-$.

\section*{Acknowledgments}
We are grateful to Kostas Orginos for providing the gauge field ensemble, which was generated using resources provided by XSEDE (supported by National Science  Foundation Grant No.~ACI-1053575).
LL was supported by the U.S. Department of Energy, Office of Science, Office of Nuclear Physics under contract DE-AC05-06OR23177.
SM and GR were supported in part by National Science Foundation Grant No.~PHY-1520996; SM and GR were also supported in part by the U.S.~Department of Energy Office of High Energy Physics under Grant No.~DE-{}SC0009913. 
SM and SS further acknowledge support by the RHIC Physics Fellow Program of the RIKEN BNL Research Center. 
JN and AP were supported in part by the U.S. Department of Energy Office of Nuclear Physics under Grant Nos.~DE{}-SC-0{}011090 and DE-{}FC02-06ER41444. 
We acknowledge funding from the  European Union's Horizon 2020 research and innovation programme under the Marie Sklodowska-Curie grant agreement No 642069. SP is a Marie Sklodowska-Curie  fellow supported by the HPC-LEAP joint doctorate program. This research used resources of the National Energy Research Scientific Computing Center (NERSC), a U.S. Department of Energy Office of Science User Facility operated under Contract No. DE-{}AC02-05CH11231.


\providecommand{\href}[2]{#2}\begingroup\raggedright\endgroup

\end{document}